\documentclass[aps,pra,reprint,10pt,amsmath,amssymb,superscriptaddress]{revtex4-1}

\usepackage[table]{xcolor}
\usepackage{hyperref,graphicx}

\newcommand{\ket}[1]{|#1\rangle}

\newcommand{\figref}[1]{Fig.~\ref{#1}}

\newcommand{\polimi}[0]{Dipartimento di Fisica - Politecnico di Milano, p.za Leonardo da Vinci 32, 20133 Milano, Italy}
\newcommand{\ifncnr}[0]{Istituto di Fotonica e Nanotecnologie - Consiglio Nazionale delle Ricerche (IFN-CNR), p.za Leonardo da Vinci 32, 20133 Milano, Italy}
\newcommand{\sapienza}[0]{Dipartimento di Fisica - Sapienza Universit\`{a} di Roma, p.le Aldo Moro 5, 00185 Roma, Italy}
\newcommand{\icfo}[0]{ICFO - Institut de Ciencies Fotoniques, The Barcelona Institute of Science and Technology, 08860 Castelldefels (Barcelona), Spain}
\newcommand{\sevilla}[0]{Departamento de F\'{i}sica Aplicada II, Universidad de Sevilla, 41012 Sevilla, Spain}

\begin{document}

\title{Single-photon quantum contextuality on a chip}

\author{Andrea Crespi}
\affiliation{\polimi}
\affiliation{\ifncnr}
\author{Marco Bentivegna}
\affiliation{\sapienza}
\author{Ioannis Pitsios}
\affiliation{\ifncnr}
\affiliation{\polimi}
\author{Davide Rusca}
\affiliation{\polimi}
\author{Davide Poderini}
\affiliation{\sapienza}
\author{Gonzalo Carvacho}
\affiliation{\sapienza}
\author{Vincenzo D'Ambrosio}
\affiliation{\sapienza}
\affiliation{\icfo}
\author{Ad\'{a}n Cabello}
\affiliation{\sevilla}
\author{Fabio Sciarrino}
\affiliation{\sapienza}
\author{Roberto Osellame}
\affiliation{\ifncnr}
\affiliation{\polimi}

\begin{abstract}
In classical physics, properties of the objects exist independently on the context, i.e. whether and how measurements are performed. Quantum physics showed this assumption to be wrong and that Nature is indeed ``contextual''. Contextuality has been observed in the simplest physical systems such as single particles, and plays fundamental roles in quantum computation advantage. Here, we demonstrate for the first time quantum contextuality in an integrated photonic chip. The chip implements different combinations of measurements on a single photon delocalized on four distinct spatial modes. We show violations of a CHSH-like non-contextuality inequality by 14 standard deviations. This paves the way to compact and portable devices for contextuality-based quantum-powered protocols.
\end{abstract}

\maketitle

\section{Introduction}

The assumption of non-contextuality, i.e., that measurements reveal properties that exist independently of whether and how measurements are carried out, lies at the heart of classical physics. The failure of this assumption in quantum theory \cite{bell1966, kochen1967} is dubbed ``contextuality'', and is a leading candidate for a notion of non-classicality with broad scope. In fact, unlike Bell nonlocality \cite{bell1964}, contextuality applies not only to space-like separated composite systems, but even to single particles. In addition, unlike macrorealism \cite{clemente2016}, the set of non-contextual correlations has a precise mathematical definition \cite{araujo2013}.
The experimental observation of contextuality can be achieved by testing correlation inequalities \cite{aspect1982,spekkens09}, which hold true whenever a non-contextual model exists, and whose violation certifies that no non-contextual model is possible. A well-established approach to test quantum contextuality is based on sequential measurements operated on a single quantum system \cite{cabello08,lapkiewicz2011, dambrosio2013,hasegawa2003,  kirchmair2009, moussa2010}. This kind of tests generally assume that measurements are sharp \cite{spekkens2014} (i.e., repeatable and minimally disturbing \cite{chiribella2014}) and that events (an event is a measurement and its outcome) have the same probability distributions in all preparation procedures \cite{spekkens2005}, even if it is possible to relax these idealizations by adopting an extended definition of non-contextuality \cite{kuajala2015}. This will be the approach followed in this work.

Many single-system quantum-contextuality-based schemes for cryptography \cite{ekert1991, spekkens09, cabello11} and randomness generation \cite{pironio2010,um2013} have been proposed in the recent years, and the mainstream interest in contextuality has skyrocketed after the proofs \cite{howard2014, delfosse2015} that it constitutes the essential resource behind the power of certain quantum computers. Quantum optics is indeed a promising approach for the realization of actual quantum computing devices \cite{Ladd10}. In particular, in the attempt to move towards scalable implementations of quantum computation and communication, a great deal of attention has been devoted to the development of integrated quantum photonics \cite{politi2008,tanzilli2012,carolan15,bentivegna15,marshall2009,sansoni2010,flamini2015} in the last decade. In fact, the need for high-fidelity operations and increasing circuital complexity \cite{carolan15, bentivegna15}, with larger number of qubits, makes the use of integrated platforms an unavoidable choice in the long term. It is therefore of prime importance to investigate whether quantum contextuality can be produced in compact and integrable devices and specifically in quantum photonic chips.

Here we perform the first on-chip test of quantum contextuality. We work with a very essential physical system, in which a single degree of freedom of a single photon, i.e. its discretized spatial position on four modes, is used to encode two qubits.
%We work with a two-qubit system: defined by four spatial modes of a single photon.  
Reconfigurable photonic circuits, realized by femtosecond laser waveguide writing, are employed both to prepare delocalized photon states across the four modes and implement different unitary operations, in order to achieve different projective measurements with the aid of single-photon detectors.

\section{A mechanical toy-model}

\begin{figure}[!t]
\centering
\includegraphics[width=\linewidth]{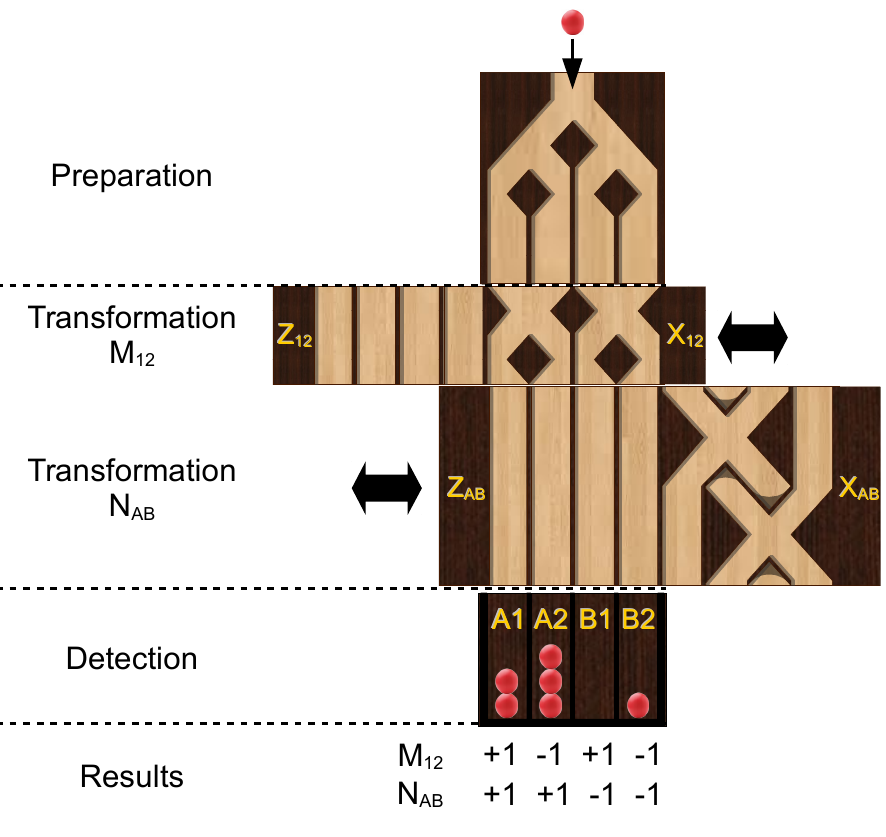}
\caption{A mechanical example: identical balls enter a modified Galton board, composed of several sections. The first section distributes the balls in the four channels according to a certain probability distribution. That is, it prepares the balls in a certain state. The second
section is reconfigurable and implements two transformations: $M_\mathrm{12}$ 
chosen between $Z_\mathrm{12}$ and $X_\mathrm{12}$, and $N_\mathrm{AB}$
chosen between $Z_\mathrm{AB}$ and $X_\mathrm{AB}$, depending on how the sliding parts are placed.
Each of these transformations, together with the detection at the bottom, constitutes a measurement 
on the distribution prepared at the first stage, whose outcome is given by the final position of the ball as indicated in the figure. Note that the measurements corresponding to $M_\mathrm{12}$ and $N_\mathrm{AB}$ are always independent in the sense that the probabilities $P(N_\mathrm{AB}=-1)$ and $P(N_\mathrm{AB}=+1)$ are independent of $M_\mathrm{12}$, and  the probabilities $P(M_\mathrm{12}=-1)$ and $P(M_\mathrm{12}=+1)$ are independent of $N_\mathrm{AB}$.}
\label{fig:classical}
\end{figure}

To get an intuitive grasp on our experiment and on its implications we shall first consider the mechanical toy-model shown in \figref{fig:classical}. This consists in a set of identical balls and a modified Galton board, composed of different sections, where the balls can be shuffled across four possible channels (A1, A2, B1, B2). The first section is a box with one input connected
to four outputs; when we throw a ball in it, it comes out at one of the four outputs, according to a certain probability distribution, which is a function of the physical characteristics of the ball and of the box. We may look at this first box as a device that \textit{prepares} the ball in a certain state. A second section of the apparatus is composed of two sliding parts that can be configured to
perform different operations. Each of the sliding parts may just let the ball fall in the same channel as it enters (the Z operations), or introduce a 50\% probability for a channel change (the X operations). $M_\mathrm{12}$ acts only on the digit, while $N_\mathrm{AB}$ acts only on the letter ($M$ and $N$ being either $Z$ or $X$). Overall, there are four possible configurations for this
second section. Balls are eventually collected at the output. We could consider the sliding sections, together with the collection stage, as an apparatus that allows to perform different measurements on the prepared state, which yield as outcome two independent bits, a letter (A, B) and a digit (1, 2).
Finally, we can conventionally assign a number (+1 or -1) to the outcomes of the two measurements, defined by the position of the two sliding parts as shown in \figref{fig:classical}.

In this classical system, the position of the ball, although only probabilistically predictable, is always defined in every moment of its evolution. The following CHSH-like non-contextuality inequality is therefore satisfied \cite{hasegawa2003} (see also Appendix~\ref{sec:derIneq}):
\begin{equation}
    S = \langle X_\mathrm{12} X_\mathrm{AB} \rangle +
    \langle X_\mathrm{12} Z_\mathrm{AB} \rangle +
    \langle Z_\mathrm{12} X_\mathrm{AB} \rangle -
    \langle Z_\mathrm{12} Z_\mathrm{AB} \rangle \leq 2,
    \label{eq:CHSH}
\end{equation}
where $\langle M_\mathrm{12} N_\mathrm{AB} \rangle$ is the average value of the product of the measurement outcomes of $M_\mathrm{12}$ and $N_\mathrm{AB}$ on a large number of events identically prepared. This inequality holds irrespectively of the specific features of the $Z$ and $X$ transformations, with the only condition that the operations implemented by the two moving parts are independent. This is intrinsically achieved since $M_\mathrm{12}$ and $N_\mathrm{AB}$ act on different bits.

\section{Integrated photonic devices}

\begin{figure*}[!tb]
\centering
\includegraphics[width=0.8\textwidth]{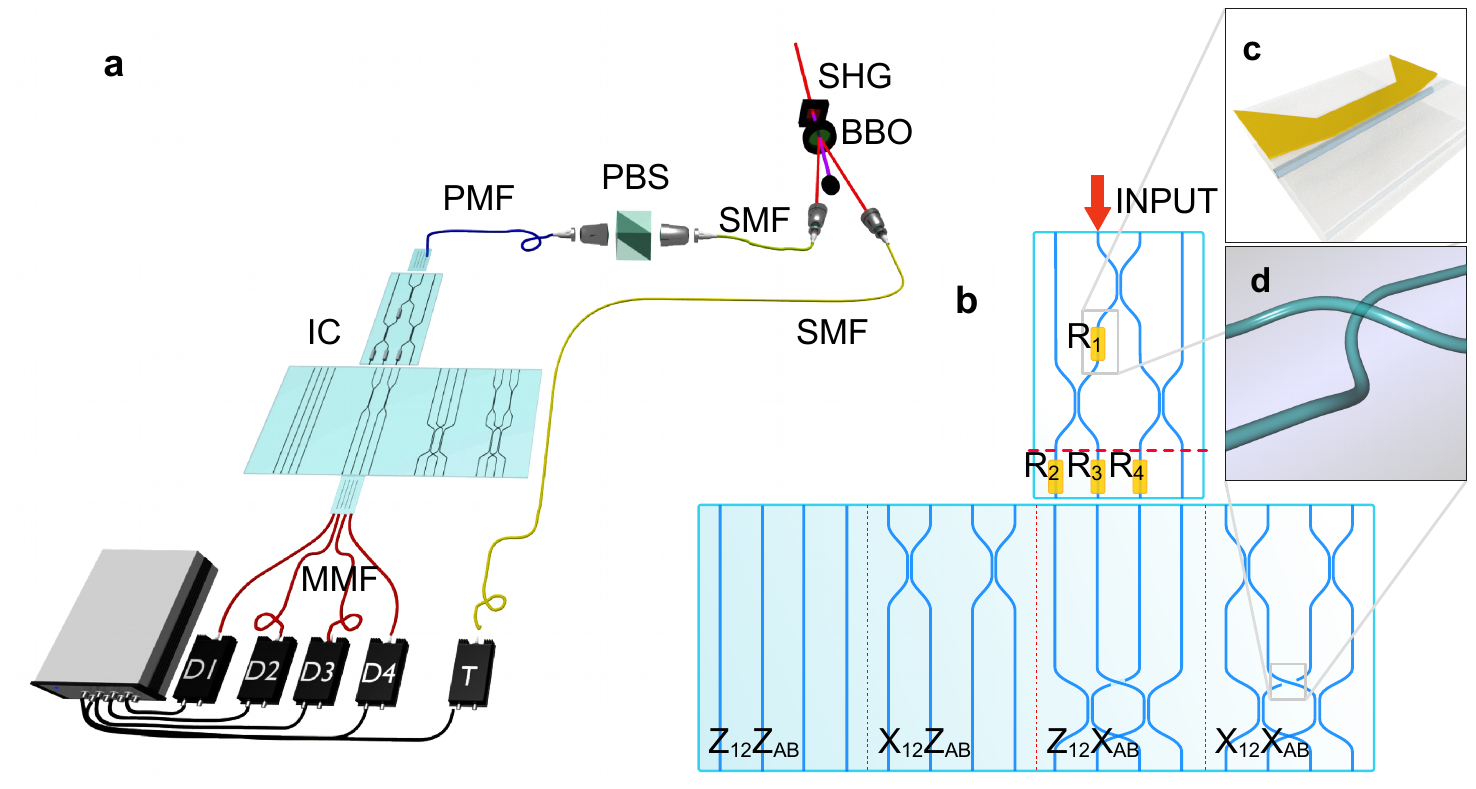}
\caption{\label{fig:experimental} Experimental setup for the contextuality measurements.
{\bf(a)} The heralded single-photon source is based on a cascade of second-harmonic-generation {by} a pulsed laser beam on a first non-linear crystal (SHG), followed by spontaneous parametric down-conversion on a BBO crystal. The generated photon pair is coupled to single
mode fibers (SMF). The trigger photon is sent directly to a detector (T), while the signal photon is first passed through a {polarizing beam splitter} (PBS) and then coupled into a polarization maintaining fiber (PMF), that injects it into the integrated photonic circuits (IC). The four outputs are coupled to single-photon detectors (D1...D4) by an array of multimode fibers (MMF). Coincidence detection of the two photons is performed by an electronic board. {\bf(b)} Detailed schematic of
the two cascaded photonic chips: the first one serves as state preparation, while the second one implements different measurements on the single-photon state. Thermo-optic phase shifters {\bf(c)} are deposited on the first chip to sweep through several different states (R1) and to calibrate the phase terms at the interface (R2, R3 and R4). The photonic circuits of the second chip exploit the three-dimensional capability of femtosecond laser waveguide writing {\bf(d)}, allowing the crossing of two waveguides without intersecting.}
\end{figure*}

The exact quantum analogue of the above classical mechanics experiment is performed by using photons instead of balls and integrated optical circuits instead of the wooden Galton board (\figref{fig:experimental}). Photons at 785~nm are provided by a heralded single-photon source, based on type-II spontaneous parametric down-conversion, which consists of a pulsed pump impinging on a BBO crystal. For each generated pair, one of the two photons acts as a trigger, while the second one is injected in a system of two cascaded integrated photonic chips, and the output is sent to single-photon detectors. Waveguides are inscribed in borosilicate glass substrate using the femtosecond laser writing technology \cite{marshall2009, sansoni2010, flamini2015} (more details about the fabrication of the integrated devices are given in Appendix~\ref{sec:wgFab}). The first chip serves as the state preparation section. The second chip, together with the detectors at the output, allows us to perform several different measurements on the state. While in our mechanical example the four different possible measurements could be implemented by adjusting two moving part (each with two allowed positions), here, for simplicity, we have fabricated four different photonic circuits, one next to the other, each implementing a different configuration. Relative translation of the second chip with respect to the first one allows to select the desired measurement.

Quantum theory provides a clear description of our photonic experiment in terms of qubits and observables. In particular, the first chip prepares single photons in a superposition state of four spatial modes, which encodes two qubits. The first qubit identifies which half of the chip is occupied
($\ket{0}$=left and $\ket{1}$=right, as the letter in the classical example) and the second gives the parity of the occupied mode ($\ket{0}$=odd and $\ket{1}$=even, as the digit in the classical example). The four states ($\ket{00}$, $\ket{01}$,$\ket{10}$ and $\ket{11}$) correspond to the states with
the photon in a well defined spatial mode. The preparation chip includes three cascaded directional couplers properly designed to produce photons in the state:
\begin{equation}
    \ket{\psi} = \frac{\ket{\mathrm{00}} e^{\iota \varphi} +
    \left(1 + \sqrt{2} \right) \left( \ket{\mathrm{01}} e^{\iota \varphi} +
    \ket{\mathrm{10}} \right) - \ket{\mathrm{11}}}{2\sqrt{2+\sqrt{2}}}
    \label{eq:state}
\end{equation}
where the term $\varphi$ can be varied by a thermo-optic phase shifter, marked as R1 in \figref{fig:experimental}b. The above photon state is defined in the circuit at the red dashed line reported in the same figure.
Three further thermo-optic shifters (R2, R3 and R4) enable a fine tuning of the optical path-lengths in the different output branches to compensate for slight geometrical misalignments when the two chips are coupled together.

The second chip, together with the fiber-coupled single-photon detectors, allows us to perform the different measurements required to evaluate the CHSH-like inequality \eqref{eq:CHSH}.  The Z and X operations are implemented respectively with straight waveguides, which let the photons proceed straight on the same modes, and balanced directional couplers, which enable mode-hopping of the photon between two modes with 50\% probability. In quantum theory, such transformations correspond nominally to the Pauli $\sigma_Z$ and $\sigma_X$ operators over the two qubits and are equivalent to basis rotations.  The Z operation leaves a qubit unchanged, so that measuring an output photon in the left or in the right mode corresponds to measuring the states $\ket{0}$ and $\ket{1}$. The X operation , which consists in the Hadamard gate, switches from the $\sigma_Z$ basis to the $\sigma_X$ one and viceversa, allowing to measure in the \{$\ket{-}$, $\ket{+}$\} basis by detecting photons in the left or right mode. By combining $\sigma_X$ and $\sigma_Z$ operators we can build the four observables $X_{12}X_{AB} = \sigma_X \otimes \sigma_X $, $X_{12}Z_{AB} = \sigma_X \otimes \sigma_Z$, $Z_{12}X_{AB}=\sigma_Z \otimes \sigma_X$  and $Z_{12}Z_{AB}=\sigma_Z \otimes \sigma_Z$, where $\sigma_j \otimes \sigma_i$ means $\sigma_i$ and $\sigma_j$ acting on the first and the second qubit, respectively.
The generic term $<M_{12}N_{AB}>$ in the inequality \eqref{eq:CHSH} ($M$ and $N$ being either $X$ or $Z$) is given by $P^{MN}_1-P^{MN}_2-P^{MN}_3+P^{MN}_4$, where $P^{MN}_i$ is the probability of finding a photon in mode $i$ after operating the transformation $M$ on the first qubit and $N$ on the second. It should be noted that the quantum operations performed by the second chip can be fully characterized using coherent light.
\begin{figure*}[tb!]
\centering
\includegraphics[width=0.7\textwidth]{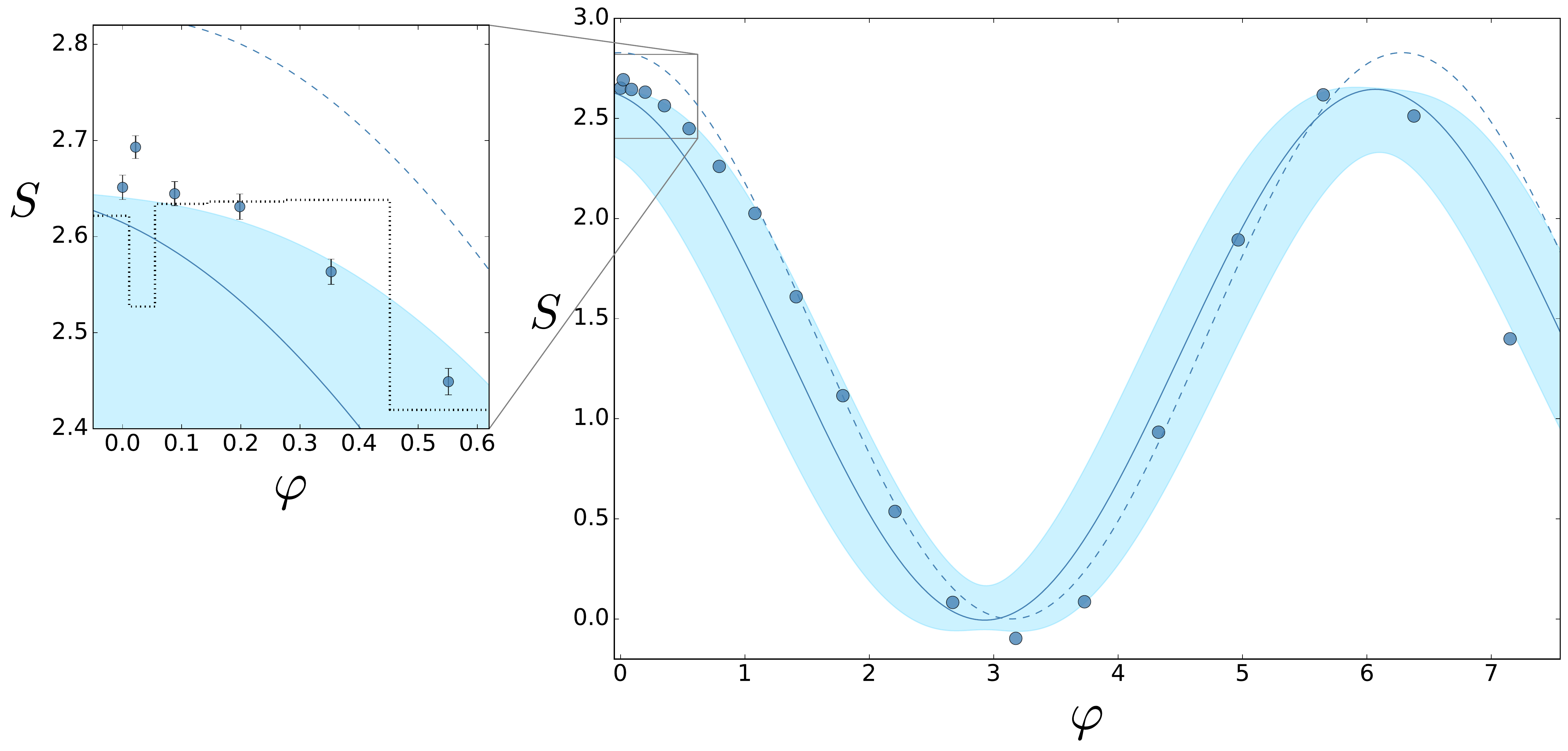}
\caption{Observed values of noncontextually-bounded quantity $S$ (left term of inequality \eqref{eq:CHSH}) as a
function of the input state phase $\varphi$. Blue points are experimental
values. Dashed line corresponds to theoretical prediction in the case of ideal
devices for state preparation and measurements, while continuous line shows the
theoretical prediction taking into account the effective trasmittivities of the
implemented devices, inferred from measurements with classical light. Blue area
shows 1$\sigma$ uncertainty region for the latter prediction, due to errors in
the phase tuning procedure. The blow-up shows a detail of the region of the graph where the maximum violation of the classical bound is observed. The dotted line represents the classical bound corrected for experimental chip imperfections as described in the Appendix~\ref{sec:nonId}.}
\label{fig:measurement}
\end{figure*}

\section{Experimental measurements}

The actual experiment is performed by collecting coincidence counts between the trigger detector and one of the output detectors for several values of dissipated power in the resistance R1 (i.e., for different phases $\varphi$ of the input state \eqref{eq:state}) and for each of the four possible measurement configurations. 
The experimental results are shown in \figref{fig:measurement} (full circles). The dashed line in the graph represents the theoretical expectation in an ideal experiment, while the continuous line is the expected trend taking into account imperfections and fabrication tolerances in the integrated photonic components (the blue shade gives a 1$\sigma$ allowance region). The experimental points are indeed well explained by this corrected model. A violation of the classical bound $S \leqslant 2$ is evident for several experimental points, around $\varphi = 0$ or $\varphi = 2 \pi$, in agreement with the predictions of quantum theory.

Recent results have shown that taking into account the experimental imperfections also affects the classical bound \cite{kuajala2015}. In particular, the inequality \eqref{eq:CHSH} should be modified to a form:
\begin{equation}
S \leq 2 + \varepsilon
\end{equation}
with $\varepsilon \geq 0$ in order to include in the classical bound the effects of non-ideal compatibility of the implemented measurements (see Appendix~\ref{sec:nonId}).
The modified classical bound is represented by the black dotted line in the blow-up of \figref{fig:measurement}. Notably, for $\varphi= 0.022$ we are able to observe experimentally a value $S=2.69 \pm 0.012$, which violates the corrected classical boundary, equal to $2.53$ for that value of $\varphi$, by $14 \sigma$.

\section{Discussion}

It is interesting to compare the behaviour of the mechanical setup of \figref{fig:classical} to the results of our experiment in integrated quantum photonics. In fact, in the first case the physical state of the ball is described not only by its position, but also by many other quantities (its shape, its speed, its orientation\ldots), which are a sort of hidden variables. Randomness there is due to ignorance of these hidden variables. A perfect knowledge of all the parameters would instead allow to predict exactly the output channel for each ball we throw in, within a classical description. 
On the contrary, in our photonic experiment, according to quantum theory, the only available degree of freedom for the photons in each point of their propagation inside the chips is their position, namely which optical mode they populate. However, even if we know precisely this information at the initial condition, i.e. in which mode the photon is injected, quantum theory would not predict exactly at which output mode the photon will exit. In fact, the occupation of any mode by the photon will remain undetermined up to the point at which it is measured. This substantial difference with the classical description is the main reason for the experimental violation of the inequality \eqref{eq:CHSH} by a quantum system, thus forbidding the existence of non-contextual hidden variables that would determine a specific trajectory for each photon in the device.

In conclusion, we have shown the first contextuality test on an integrated photonic chip, demonstrating the reliability and versatility of current photonic integration techniques for testing quantum properties, and for producing compact and portable devices capable to exploit and certify the enhanced capabilities of quantum technologies. In perspective, this technology could be used to implement sources of correlations with computational power \cite{anders2009} integrable within conventional hardware.

We highlight that the intrinsic stability of integrated waveguide circuits has allowed us to design and perform an experiment involving  only the spatial degree of freedom of a single photon, and in particular based only on interference between different paths. Our experimental setup thus makes it easy to visualize that contextuality is a fundamental property of quantum systems that is a direct consequence of wavefunction interference. 

\appendix

\section{Validity of the CHSH inequality}
\label{sec:derIneq}

The CHSH-like inequality
\begin{equation}
\langle X_\mathrm{12} X_\mathrm{AB} \rangle + \langle X_\mathrm{12}
Z_\mathrm{AB} \rangle + \langle Z_\mathrm{12} X_\mathrm{AB} \rangle - \langle
Z_\mathrm{12} Z_\mathrm{AB} \rangle \leq 2
\label{eq:CHSHlike}
\end{equation}
holds true on three fundamental assumptions:
\begin{itemize}
    \item \emph{Realism}: The outcomes of a measurement are determined \emph{before}
        the actual measurement.
    \item \emph{Non-contextuality}: The outcome of a measurement does not depend on
        which others \emph{compatible} measurment(s) are simultaneously performed.
    \item \emph{Compatibility}: The four couples of observables 
        \begin{align}
            &(X_\mathrm{12}, X_\mathrm{AB})
            &(X_\mathrm{12}, Z_\mathrm{AB}) \notag
            \\
            &(Z_\mathrm{12}, X_\mathrm{AB})
            &(Z_\mathrm{12}, Z_\mathrm{AB}) \notag
            \label{eq:comp_rel}
        \end{align}
        are compatible. 
\end{itemize}
The notion of compatibility outside of the framework of quantum mechanics needs clarification. Here we call two measurements \emph{compatible} when they can be measured simultaneously without any disturbance.

If the above assumptions are satisfied, knowing that the measurement outcomes of the observables can only take the values $\pm 1$, it is easy to see that the left hand side of \eqref{eq:CHSHlike} can never exceed $2$.
Therefore, if a violation of the inequality is experimentally observed, it follows that one of the above assumptions is not satisfied. In particular our experiment aims at disproving the combination of the first two, called \emph{non-contextual realism}, by ensuring that the third holds true. This means that the measurement of $M_\mathrm{12}$ (or $N_\mathrm{AB}$) made jointly with $Z_\mathrm{AB}$ (or $Z_\mathrm{12}$) should yield the same result as the one made with ${X}_\mathrm{AB}$ (or ${X}_\mathrm{12}$), for every input state.

\section{Waveguide fabrication}
\label{sec:wgFab}

Waveguides were fabricated by direct femtosecond laser writing using a Yb:KYW cavity-dumped mode-locked oscillator ($\lambda$~=~1030~nm). Ultrafast pulses (300~fs pulse duration, 1~MHz repetition rate) were focused using a microscope objective 0.6~NA, 50$\times$, into the transparent volume of an alumino-borosilicate glass (Corning, EAGLE 2000), producing a local and permanent refractive index increase. Translation of the sample with a constant tangential velocity of 40~mm~$s^{-1}$ (Aerotech FiberGLIDE 3D air-bearing stages), allows to draw the desired waveguiding paths. In the state-preparation chip waveguides were inscribed at 25~$\mu$m depth, with 220~nJ pulse energy. In the measurement chip waveguides were inscribed at 70~$\mu$m depth and 230~nJ pulse energy. The size of the two chips is respectively 49~mm~$\times$~24~mm and 65~mm~$\times$~27~mm. The thermo-optic phase shifters are fabricated by depositing a thin gold layer on the top surface of the chip and by patterning the resistors by laser ablation, with the same femtosecond laser source used for the waveguide fabrication (according to the method described in Ref.~\cite{flamini2015}).

\section{Non-ideal compatibility}
\label{sec:nonId}

In a real situation the compatibility relations discussed in Appendix~\ref{sec:derIneq}
might be not always satisfied. In our specific case this may due to the fact that we are using different devices to measure the same property in different contexts.

As explained in Ref.~\cite{kuajala2015} in this case it is possible to derive a different bound for the inequality, whose violation means that the contextuality of the system cannot be explained in terms of the ``trivial'' contextuality given by the lack of ideal compatibility.
Namely, the new bound for \eqref{eq:CHSHlike} becomes:
\begin{align}
	&S \le 2 + \varepsilon  \notag \\
    &\varepsilon = \sum_M \left| \langle{M^X}\rangle - \langle{M^Z}\rangle\right|
\end{align}
where we have introduced the notation $M^N$ to distinguish the measurement $M$ performed simultaneously with $N$, and the sum is extended to the four measurements $X_\mathrm{12}, X_\mathrm{AB}, Z_\mathrm{12}, Z_\mathrm{AB}$. Note that the values of different $\langle{M^N}\rangle$ can be retrieved from the same set of experimental data used to evaluate \eqref{eq:CHSHlike}.

\vspace{1em}

\textbf{Acknowledgements}
This work was financially supported by the H2020-FETPROACT-2014 Grant QUCHIP (Quantum Simulation on a Photonic Chip;
grant agreement no. 641039): http://www.quchip.eu, and by the Marie Curie Initial Training Network PICQUE (Photonic Integrated Compound Quantum Encoding, grant agreement no. 608062, FP7-PEOPLE-2013-ITN, http://www.picque.eu).  A.Ca. acknowledges support from Project No. FIS2014-60843-P, ``Advanced Quantum Information'' (MINECO, Spain), with FEDER funds and the project ``Photonic Quantum Information'' (Knut and Alice Wallenberg Foundation, Sweden). 

\textbf{Author contributions}
A.Ca., F.S. and R.O. conceived the experiment. A.Cr., I.P., D.R. and R.O. designed and fabricated the integrated photonic devices and performed classical characterization. M.B., D.P., G.C., V.D'A. and F.S. performed the experiments with single photons and analyzed the data. All authors discussed the results and contributed to the writing of the paper.

\end{document}